\pdfoutput=1
\documentclass [runningheads]{llncs}
\usepackage{graphicx}
\usepackage{amsmath}
\usepackage{amssymb}
\usepackage{multirow}
\usepackage{algorithm}  
\usepackage{algpseudocode}  
\usepackage{caption2}
\usepackage{subfigure}
\usepackage{makecell}
\makeatletter
\newenvironment{breakablealgorithm}
  {
   \begin{center}
     \refstepcounter{algorithm}
     \hrule height.8pt depth0pt \kern2pt
     \renewcommand{\caption} [2] [\relax]{
       {\raggedright\textbf{\ALG@name~\thealgorithm} ##2\par}%
       \ifx\relax##1\relax 
         \addcontentsline{loa}{algorithm}{\protect\numberline{\thealgorithm}##2}%
       \else 
         \addcontentsline{loa}{algorithm}{\protect\numberline{\thealgorithm}##1}%
       \fi
       \kern2pt\hrule\kern2pt
     }
  }{
     \kern2pt\hrule\relax
   \end{center}
  }
\makeatother

\begin{document}
\title{Improving Robustness of TCM-based Robust Steganography with Variable Robustness Cost}
%
%
\author{Jimin Zhang\inst{1,2} \and 
Xianfeng Zhao\inst{1,2} \and
Xiaolei He\inst{1,2}}
\authorrunning{J Zhang, X Zhao, et al.}
%
\institute{State Key Laboratory of Information Security, Institute of Information Engineering, Chinese Academy of Sciences, Beijing 100195, China \and
School of Cyber Security, University of Chinese Academy of Sciences, Beijing 100195, China \\
    \email{zhangjimin@iie.ac.cn, zhaoxianfeng@iie.ac.cn, hexiaolei@iie.ac.cn}}

\maketitle            
\begin{abstract}
Recent study has found out that after multiple times of recompression, the DCT coefficients of JPEG image can form an embedding domain that is robust to recompression, which is called transport channel matching (TCM) method. Because the cost function of the adaptive steganography does not consider the impact of modification on the robustness,  the modified DCT coefficients of the stego image after TCM will change after recompression. To reduce the number of changed coefficients after recompression, this paper proposes a robust steganography algorithm which dynamically updates the robustness cost of every DCT coefficient. The robustness cost proposed is calculated by testing whether the modified DCT coefficient can resist recompression in every step of STC embedding process. By adding robustness cost to the distortion cost and using the framework of STC embedding algorithm to embed the message, the stego images have good performance both in robustness and security. The experimental results show that the proposed algorithm can significantly enhance the robustness of stego images, and the embedded messages could be extracted correctly at almost all cases when recompressing with a lower quality factor and recompression process is known to the user of proposed algorithm.

\end{abstract}
\section{Introduction}
Steganography is the technology that embeds secret message in the innocent-looking multimedia file such as image, video, audio or text. Covert communication could be achieved using steganography in the public channel. As a kind of public channel, social networks could improve the security of covert communication for anyone can download the stego image so that the receiver could be more difficult to find. But social networks usually perform JPEG recompression to the images uploaded by users with fixed quantization table. Because the JPEG recompression is a lossy process, the DCT coefficients of uploaded image could be changed so that the information hidden in the uploaded image using adaptive steganography like J-UNIWARAD \cite{holub2013digital}, UERD \cite{guo2015using} algorithm could not be extracted correctly, which increases the difficulty of transmitting secret information in the public channel using steganography.                                              

To achieve robust embedding in JPEG recompression channels, some robust steganography algorithms have been developed in recent years. Zhang et al. \cite{zhang2015jpeg,zhang2017joint,zhang2018dither}  propose three algorithms that utilize the embedding domain of watermarking and design distortion functions to achieve safe and robust embedding. Zhao et al. \cite{zhao2018improving} propose the transport channel matching (TCM) method that uploads images to the social networks several times to obtain the images whose coefficients could not be changed anymore by uploading, and adaptive steganography is used to embed information in the channel matched image. Kin-Cleaves et al. \cite{kin2018adaptive} propose the DSTC algorithm which uses two embedding matrices to correct stego image and embed secret message respectively. Tao et al. \cite{tao2018towards} propose a robust steganography which utilizes the embedded channel image to guide the change step of the cover image. 

When the recompression process is known to the user of steganography, transport channel matching method could be used to generate robust images with local simulation of recompression process in social networks.  The TCM-based  robust steganography using distortion cost which calculates the cost of every DCT coefficient based on the influence of embedding to the image content, but the effect on the robust domain of recompression is not related, resulting in many changes in stego images after channel transmitting. Error correction code could be used to correct errors. When errors are large, error coding with low embedding efficiency can be used to correctly restore the hidden message, which increases the embedding impact on the stego image. 

To add the influence of embedding to the robust domain, a robustness cost is proposed in this paper. By combining the robustness cost and distortion cost, the cost  used in embedding algorithm proposed can reflect embedding impact both on image content and robustness. The stego image embedded with STC embedding algorithm \cite{filler2011minimizing} has good performance both in security and robustness. 

The structure of this paper is organized as follows. In section 2 the TCM-based robust steganography and the framework of calculating variable cost are introduced. In section 3 the proposed robust embedding method is described in detail. In section 4 some experimental results are shown. Section 5 is the conclusion of this paper.
\section{Related Works}
\subsection{TCM-based robust steganography}
The execution process of TCM-based robust steganography algorithm could be divided into two steps. Firstly, by uploading image to social networks multiple times, a robust image that few or no coefficients will change after being uploaded to the social networks is generated. Secondly, using adaptive steganography algorithm to embed message in the  image after TCM which ensures the security performance. The process of recompression of social networks could be described as follows. Assuming that the DCT coefficients of a $8 \times 8$ DCT block in JPEG image that need to be uploaded is $ \vec {D}_{q1} $, firstly the coefficients are decompressed into the spatial domain which is described as:
\begin{equation}\label{dct2spatial}
\begin{aligned}
&\vec{C}_{q1} = \vec{q}_{1} \times \vec{D}_{q1} \\
&\vec{x}_{b}^{\prime} = \operatorname{IDCT}(\vec{C}_{q1}) \\
&\vec{x}_{b} =  [\vec{x}_{b}^{\prime}] \\
&\vec{x}_{b+} = \vec{x}_{b} + 128,
\end{aligned}
\end{equation}
where $\vec{q}_{1}$ represents the quantization table of JPEG images, $\operatorname{IDCT}(\cdot)$ represents the inverse DCT transform, $ [\cdot]$ represents the rounding operation, and $\times$ represents the multiplication of corresponding  elements in two matrices. Then the process of compressing $\vec{x}_{b+}$ to DCT domain is described as:
\begin{equation}\label{spatial2dct}
\begin{aligned}
&\vec{x}_{b-} = \vec{x}_{b+}-128 \\
&\vec{C}_{q2} = \operatorname{DCT}(\vec{b-}) \\
&\vec{D}_{q2}^{\prime} = \frac{\vec{C}_{q2}}{\vec{q}_{2}} \\
&\vec{D}_{q2} =  [\vec{D}_{q2}^{\prime}],
\end{aligned}
\end{equation}
where $DCT(\cdot)$ is the DCT transform, $\vec{q}_{2}$ is the quantization table of the channel, and the division is the division of the corresponding positions of matrices.
When $\vec{q}_{1}$ is equal to $\vec{q}_{2}$, the difference between $\vec{D}_{q1}$ and $\vec{D}_{q2}$ is due to the spatial rounding operation in (\ref{dct2spatial}). After second recompression, the errors of DCT coefficients caused by spatial rounding operation will be smaller than the errors of DCT coefficients after first recompression. When the image is recompressed multiple times, the errors of DCT coefficients before and after compression caused by spatial rounding will become smaller and smaller. Until there is no change, or the changed number does not reduce any more, transport channel matched image which is robust to the recompression is acquired. The adaptive steganography could be used then to embed secret message in the image after TCM, and the entire embedding algorithm is called as JCRISBE algorithm \cite{zhao2018improving}.
\subsection{Variable embedding cost}
The embedding process of the adaptive steganography algorithm can be divided into two parts: the calculation of the cost function and the embedding process using STC algorithm. For computational convenience, adaptive steganography cost is calculated based on the additive model, which assumes that the interference from embedded points does not affect each other. Taking UNIWARD cost calculation as an example, in order to distinguish between the texture area and the smooth area, three wavelet filters in different directions are used. The filtered images corresponding to first-level wavelet LH, HL and HH decomposed components are represented as $\vec{W}^{(1)} $, $\vec{W}^{(2)} $and $\vec{W}^{(3)} $.  For the cover image $\vec{X}$ whose size is $n_{1} \times n_{2}$, the distortion of stego image $\vec{Y}$ is defined as:
\begin{equation}\label{uniward_1}
D(\vec{X},\vec{Y})\triangleq\sum_{k=1}^{3}\sum_{u=1}^{n_{1}}\sum_{v=1}^{n_{2}}\frac{\left|W_{uv}^{(k)}(\vec{X})-W_{uv}^{(k)}(\vec{Y})\right|}{\sigma+\left| W_{uv}^{(k)}(\vec{X})\right|},
\end{equation}
where $\sigma>0$ is a constant that prevents the denominator from being zero. To represent the additive model of UNIWARD, defining the stego image generated by changing the $ij$th pixel $X_{ij}$ of the original image as $Y_{ij}$, and the cost by changing this pixel is described as:
\begin{equation}\label{uniward_2}
\rho_{ij}(\vec{X},Y_{ij})\triangleq D(\vec{X},\vec{X_{\sim ij}}Y_{ij}),
\end{equation}
where $\vec{X_{\sim ij}}Y_{ij}$ means only $X_{ij}$ changed into $Y_{ij}$ in the cover image. The UNIWARD cost of the additive model can be expressed as:
\begin{equation}\label{uniward_3}
D_{A}(\vec{X},\vec{Y}) = \sum_{i=1}^{n_{1}} \sum_{j=1}^{n2} \rho_{ij} (\vec{X} ,Y_{ij}) [X_{ij}\neq Y_{ij}].
\end{equation}
Since the additive model assumes that each embedded pixel does not interfere with each other, the calculation process can be simplified. From the experimental results of the non-additive steganography algorithm \cite{7109899}, it can be seen that if the embedding direction of other points around the embedding point is taken into consideration, the security performance can be improved. Since the current non-additive embedding model \cite{filler2010gibbs} based on Gibbs embedding has high computing complexity level, \cite{pevny2018exploring} proposed a non-additive embedding model by calculating the variable cost at every embedding process of STC. This method saves the embedded image in each process of STC embedding. When starting to calculate the UNIWARD cost of each state in STC embedding process, the modified UNIWARD cost based on entire stego image is calculated. Since the optimal modification path of each step is selected based on saved temporary stego image, the final embedding result is not selected based on the original cover image. Because the stego image can correctly extract the embedded information, and the influence of other modified pixels to the embedding cost is considered during the embedding process,  it can achieve non-additive steganography based on STC embedding algorithm.

\section{Proposed Method}
\begin{figure} [!htbp]
\centering
\subfigure []{
\label{fig:subfig:a} 
\includegraphics [width=0.31\textwidth]{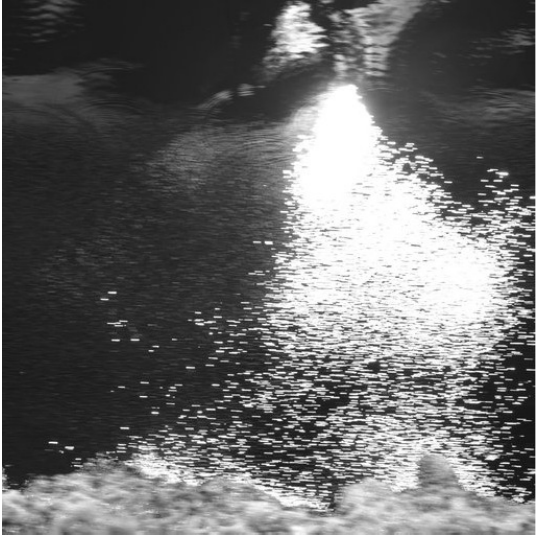}}
\subfigure []{
\label{fig:subfig:b} 
\includegraphics [width=0.31\textwidth]{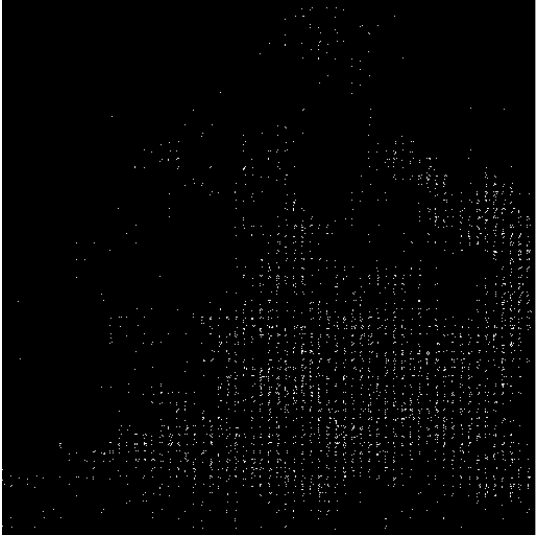}}
\subfigure []{
\label{fig:subfig:c} 
\includegraphics [width=0.31\textwidth]{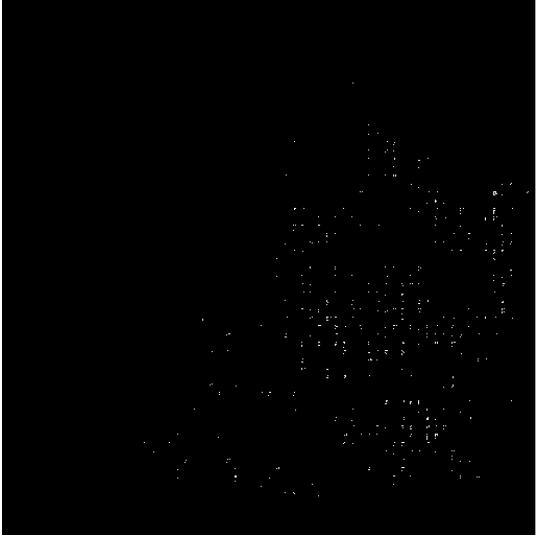}}
\caption{(a)77.jpg generated from BossBase database \cite{bas2011break}, (b)Changed coefficients using J-UNIWARD steganography, (c)Changed coefficients between stego image and JPEG recompressed stego image.}
\label{fig:subfig} 
\end{figure}

After transport channel matching, the DCT coefficients of the JPEG image will not change or only small number of coefficients will change when they are uploaded to the recompression channel without being modified by steganography algorithm. When the image is embedded the interference caused by the embedding is introduced, the stable state of DCT coefficients formed by channel matching method will be broken. In order to demonstrate how the embedding operation breaks the robust domain, a JPEG image with quality factor 85 is generated from BossBase database \cite{bas2011break} shown as \ref{fig:subfig:a}. The changed coefficients of this image using J-UNIWARD cost and binary STC embedding at payload 0.2 bpnzac is shown in \ref{fig:subfig:b}, and the changed coefficients after recompression is shown in \ref{fig:subfig:c}.

From figure \ref{fig:subfig}, it can be seen that the changing possibility of DCT coefficients after recompression is related to the position the embedding happened. The embedding coefficients in the high-frequency regions are less resistant to recompression than those in the low-frequency regions. In the other hand, by analyzing the impact of the modified coefficients on the image robustness, the recompression recoverability of a certain DCT coefficient is related to the modification of other DCT coefficients in the same DCT block. When there are many modified coefficients in a same DCT block, a new changed coefficient will cause multiple coefficients in the same DCT block to change after recompression. Therefore, when checking the recoverability of a modified coefficient after recompression, the interference of other modified coefficients in the same DCT block should be considered. At this point, the non-additive embedding model is suitable for the robust steganography algorithm. Based on the above analysis, we propose a robust steganography algorithm called Robust Steganography with Variable Robustness Cost (RSVRC), which dynamically updates the robustness cost.

The distortion cost reflects the impact of embedding operation on the image content. The changed coefficients based on distortion cost gathers in high-frequency regions. These regions are difficult to model thus having high security. However, the impact of embedding operation on the robustness against recompression is not considered when distortion cost is used to achieve adaptive steganography. In order to improve the ability of embedded points to resist recompression,  embedding algorithm should select locations with both high security and robustness. The algorithm proposed is based on the J-UNIWARD embedding cost and the robustness cost calculated by the non-additive embedding model proposed by Pevny et al. \cite{pevny2018exploring}. Different from the process of dynamically modifying the distortion cost proposed by  \cite{pevny2018exploring}, the proposed algorithm does not modify the distortion cost during the embedding process, but only updates the robustness cost by checking whether the modified coefficient can resist recompression. The calculation of the robustness cost  is defined as follows.

Assuming the coefficient $X_{ij}$ is located in the $8 \times 8$ DCT block $\vec{X}$, modification of $X_{ij}$ is achieved by randomly plus 1 or minus 1 using binary STC embedding algorithm. The DCT block after modifying the coefficient $X_{ij}$ is marked as $\vec{Y}$, and the DCT block after recompressing  $\vec{Y}$  calculated using (\ref{dct2spatial})  and (\ref{spatial2dct}) is marked as $\vec{Y^{'}}$. Then the robustness cost of modifying $X_{ij}$ is calculated as:
\begin{equation} \label{eq:rij}
r_{ij}=\left\{
\begin{aligned}
& 0 & \vec{Y} = \vec{Y^{'}}\\
& C & \vec{Y} \neq \vec{Y^{'}},
\end{aligned}
\right.
\end{equation}
where $C$ is a constant value representing the cost when changing a non-robust coefficient in the embedding process. The robustness cost of not modifying $X_{ij}$ is set as 0, which based on the hypothesis that the robust domain does not change when the coefficient is not modified. If there are errors in the DCT block after recompression due to other previous modified coefficients instead of modified $X_{ij}$, the cost of modifying $X_{ij}$ is also set as $C$ according to (\ref{eq:rij}), which aims to delete the embedding path which modifies coefficients in this DCT block in the STC embedding process by introducing more cost. The constant $C$ is set as the maximum value of the distortion cost which assumes that changing a non-robust point is as same as changing the coefficient whose distortion cost is high. 

\begin{figure}
\centering
\includegraphics [width=0.8\textwidth]{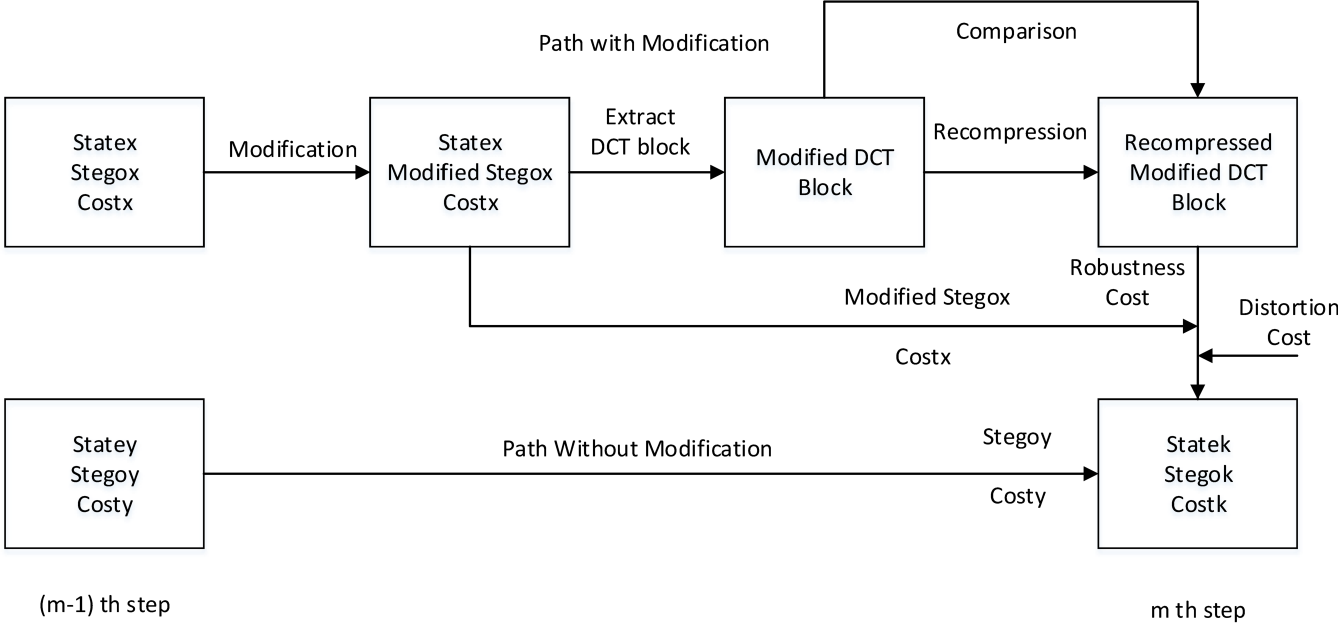}
\caption{The process of choosing state transition path of RSVRC.} \label{figrsvrc}
\end{figure}

When RSVRC embeds message using STC algorithm, the process of choosing the state transition path is described in figure \ref{figrsvrc}. Supposing that the state being tested is the $k$th state of $m$th step in the STC embedding process, the state has two transition paths into it, which correspond to the transition path with modified coefficient and the transition path without modifying coefficient. The cost of transition path with modified coefficient is the sum of the distortion cost and robustness cost introduced by modifying the coefficient, which is set to $\rho_{ij} + r_{ij}$. The cost of transition path without modifying coefficient is equal to 0. The path with smaller cost
calculated as the sum of the cost of starting state and the cost of transition path  is selected, and the image in the testing state is set according to the transition path and image saved in starting state of selected path. After the state transition path of the entire image is established, the state that matches the last embedding message and has the minimum cost can be found in the last step of embedding, and the image saved in this state is the stego image. In order to be different from the traditional STC embedding algorithm and Variable-Cost STC \cite{pevny2018exploring}, the embedding process proposed is named VRCSTC (Variable Robustness Cost STC). Algorithm \ref{alg:vrcstcs} describes the execution process of VRCSTC algorithm.

\begin{breakablealgorithm}
\label{alg:vrcstcs}
 \caption{VRCSTC Algorithm} 
 \begin{algorithmic} [1] 
  \State $C \gets max(rho)$
  \State $wght [1] \gets 0$
  \State $wght [2,2^{h}] \gets infinity$
  \State $img\_list [1] \gets X$
  \State $img\_list [2,2^{h}] \gets  []$ //set as empty
  \State $indx \gets 1$
  \For{each $i\in  [1,length(msg)]$} //for each block 
  	\For{each $j\in  [1,width]$}  //for each j in width of H\_hat 
   		\For{each $k\in  [0,2^{h}-1]$}
   		  \State $pre\_state \gets k \  xor \  H\_hat [j]$
   		  \State $w0 \gets wght [k+1] + X [indx]*rho [indx]$
   		  \State $pre\_stego\_0 \gets img\_list [k+1]$
   		  \If{ $w0 \neq infinity$}
   		  	\If { $X [indx] = 1$}
	      		\State randomly set change as +1 or -1
	          \State $pre\_stego\_0 [indx] \gets pre\_stego\_0 [indx] + change$
	          \State $test\_block \gets get\_test\_block(pre\_stego\_0 ,indx) $ 
			      \State $robustness\_cost \gets get\_robustness\_cost(test\_block,C)$
	          \State $w0 \gets w0 + robustness\_cost$
	       \EndIf
	      \EndIf
	      \State $w1 \gets wght [pre\_state+1]+(1-X [indx])*rho [indx]$
        \State $pre\_stego\_1 \gets img\_list [pre\_state+1]$
        \If{$w1 \neq infinity$}
        	\If{$X [indx] = 0$}
        		\State randomly set change as +1 or -1
	       		\State  $pre\_stego\_1  [indx] \gets pre\_stego\_1  [indx] + change$
	       		\State $test\_block \gets get\_test\_block(pre\_stego\_1,indx)$
	       		\State $robustness\_cost \gets get\_robustness\_cost(test\_block,C)$
	       	  \State $ w1 \gets w1 + robustness\_cost$
	       \EndIf
	     \EndIf
	     \If {$w1 < w0$}
         \State 	$new\_img\_list [k+1] \gets pre\_stego1$
       \ElsIf{$w1 = w0$}
         \If{$X [indx] = 1$}
         		\State  $new\_img\_list [k+1] \gets pre\_stego\_1$
     		  \Else 
         		\State  $new\_img\_list [k+1] \gets pre\_stego\_0$
     		  \EndIf
      \Else
         \State    $new\_img\_list [k+1] \gets pre\_stego\_0$
      \EndIf
      \State $new\_wght [k+1] \gets min(w0,w1)$ 
	  \EndFor
	  \State $indx \gets indx + 1$
	  \State $wght [1,2^{h}] \gets new\_wght [1,2^{h}]$
	  \State $img\_list [1,2^{h}] \gets new\_img\_list [1,2^{h}]$ 
	 \EndFor
   \For{each $j\in  [0,2^{h-1}-1]$}
  				\State $wght [j+1] \gets wght [2*j+msg(i)+1]$
        \State $img\_list [j+1] \gets img\_list [2*j+msg(i)+1]$
   \EndFor
   \For{each $ j \in  [2^{h-1}+1,2^h]$}
 	     \State $wght [j] \gets infinity$;
 	     \State $img\_list [j] \gets  []$
    \EndFor
  \EndFor
  \State $min\_cost\_index \gets get\_min\_index(wght)$
	\State $stego \gets img\_list [min\_cost\_index]$
  \label{code:recentEnd} 
 \end{algorithmic} 
\end{breakablealgorithm} 

The embedding process of the RSVRC algorithm based on TCM is described in the algorithm \ref{alg:rskrqembed}.
 \begin{algorithm} [htb]
 \caption{The embedding process of the RSVRC algorithm}
 \label{alg:rskrqembed}
 \begin{algorithmic} [1]
  \Require
   The file name of cover;
   The secret information $m_{s}$;
   Random permutation key;
   The quality factor $q$ of the recompression channel;
   BCH coding parameter;
   Recompression time threshold $C_{T}$ of transport channel matching algorithm.
  \Ensure
   Stego image;
  \State compress cover image multiple times based on the quality factor $q$ of the recompression channel, and stop when the channel error rate caused by the recompression is 0 or the number of times of recompression  reaches the threshold $C_{T}$;
  \State use BCH error correction coding to encode the secret information $m_{s}$ according to the BCH coding parameter;
  \State permute error correction coded message randomly using  random permutation key;
  \State use VRCSTC algorithm to embed message, generate the stego image;\\
  \Return stego.
 \end{algorithmic}
\end{algorithm}

\section{Experiment}
\subsection{Experimental setup}
The experiment using BossBase-1.0.1-cover \cite{bas2011break} database as the cover images, which contains 10,000 grayscale spatial images with the size of 512$\times$512. In order to test the ability of JPEG images to resist recompression, the imread and imwrite methods in MATLAB are used to generate JEPG images with quality factors of 75, 85, and 95, respectively.

The robust steganography algorithm compared with the algorithm proposed is JCRISBE algorithm \cite{zhao2018improving}. The distortion cost used in the embedding algorithm is J-UNIWARD. The maximum recompression time $C_{T} $ used in the experiment is 12. Due to the high complexity of the algorithm, in order to reduce the embedding time, the height of the small matrix  $h$ in STC is 3. In order to improve the efficiency of the algorithm, the image of 512$\times$512 was split into 16 subimages of 128$\times$128 based on randomly selecting $8 \times 8$ DCT blocks, and the message is equally divided into 16 parts when embedding with RSVRC and JCRISBE using matlab parallel computing toolbox. 

In order to compare the performance of different algorithms, channel error rate $P_{e}$, information bit error rate $P_{s}$, successful extraction rate $R_{s}$ and coding efficiency $e$ are used to evaluate the performance of algorithms in the experiment. The channel error rate $P_{e}$ is equal to the ratio of the number of changed coefficients after recompression to the number of DCT coefficients in the stego image. The information bit error rate $P_{s}$ is equal to the number of error information bits extracted using STC extraction algorithm to the total number of embedded bits. The successful extraction rate $R_{s}$ is equal to the number of images which correctly restore the message after recompression to the number of images used in the experiment. The coding efficiency $e$ is equal to the ratio of the message bits to the length of coding block of BCH coding. 
\subsection{The influence of RSVRC to the robustness}
\begin{figure} [!htbp]
\centering
\subfigure []{
\label{fig2:subfig:a} 
\includegraphics [width=0.4\textwidth]{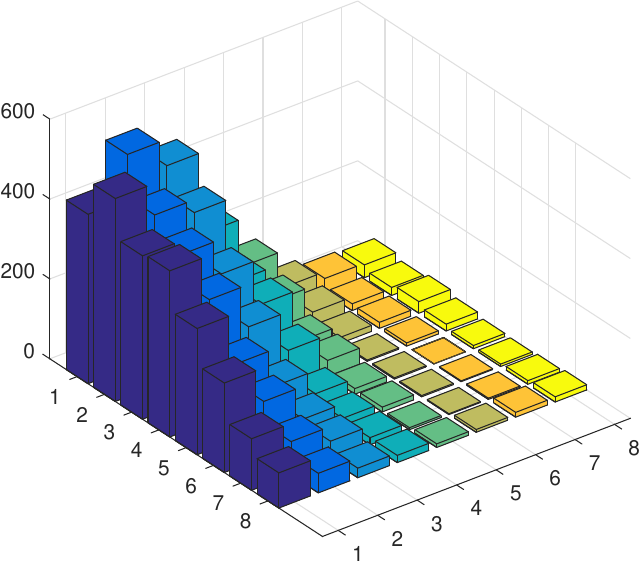}}
\hspace{0.5in}
\subfigure []{
\label{fig2:subfig:b} 
\includegraphics [width=0.4\textwidth]{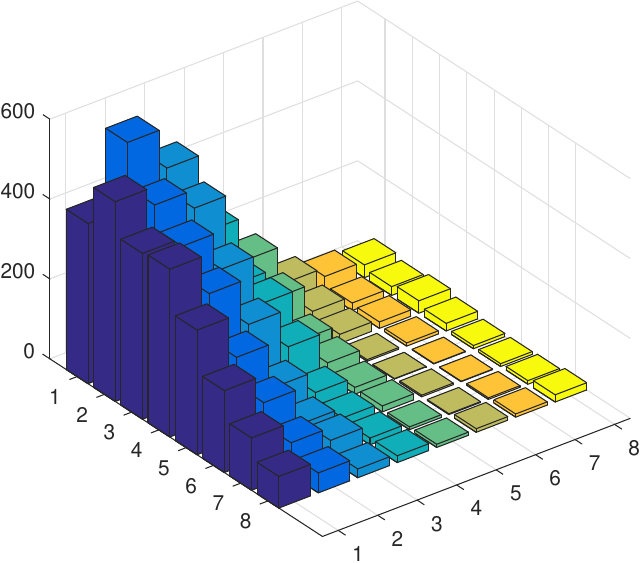}}
\caption{(a)Number of changed DCT coefficients in different DCT modes when embedding using JCRISBE, (b)Number of changed DCT coefficients in different DCT modes when embedding using RSVRC.}
\label{fig2:subfig} 
\end{figure}

In order to compare the impact of the steganography algorithm on the robustness, this part uses the JCRISBE algorithm and the RSVRC algorithm to embed information in the cover image and analyzes the embedding position and recompression error in different DCT modes. The experiment firstly compares the modification number introduced by embedding using RSVRC and JCRISBE at different DCT modes. The number of modifications of different embedding algorithms is shown in figure \ref{fig2:subfig}.

It can be seen from  figure \ref{fig2:subfig} that the number of modifications in different DCT modes of two algorithms is similar, which shows that when embedding with dynamically updated cost, the RSVRC algorithm does not embed information in the high frequency modes in the DCT block which have good robustness performance. In order to ensure the security of steganography, the adaptive steganography algorithm usually embeds message by modifying the DCT coefficients in the mid- and low-frequency modes. Therefore, the RSVRC steganography algorithm modifies DCT coefficients in the mid- and low-frequency modes to maintain the security performance.

\begin{figure} [!htbp]
\centering
\subfigure []{
\label{fig3:subfig:a} 
\includegraphics [width=0.4\textwidth]{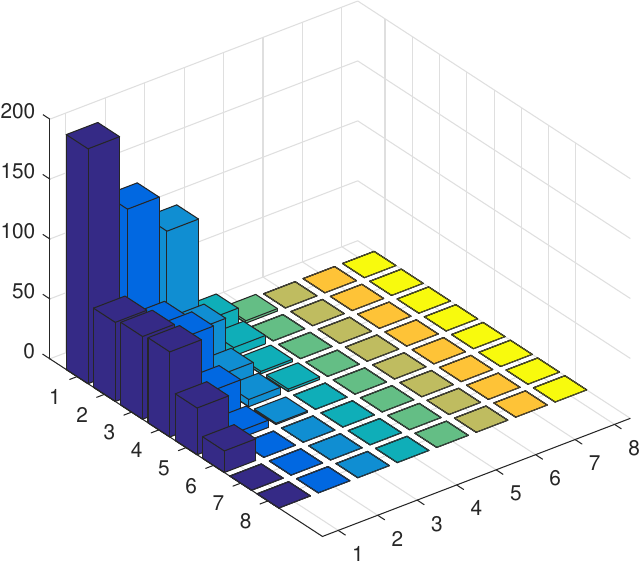}}
\hspace{0.5in}
\subfigure []{
\label{fig3:subfig:b} 
\includegraphics [width=0.4\textwidth]{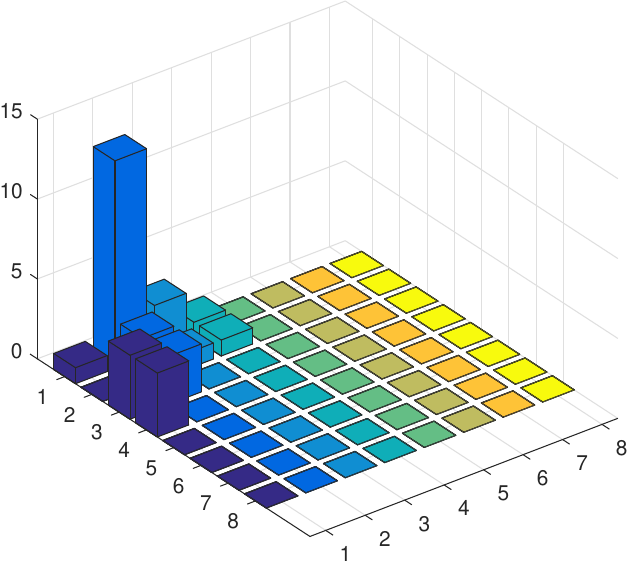}}
\caption{(a)Number of recompression errors in different DCT modes using JCRISBE,  (b)Number of recompression errors in different DCT modes using RSVRC.}
\label{fig3:subfig} 
\end{figure}

After the embedded image is compressed by recompression channel, the number of changed DCT coefficients caused by recompression in different DCT modes is shown in figure \ref{fig3:subfig}. JCRISBE algorithm and RSVRC algorithm have similar number of modifications in different modes during embedding, but the number of recompression errors in different modes is quite different. This is because the RSVRC algorithm considers robustness by introducing the robustness cost, which encourages the STC embedding algorithm to preferentially select the path with lower distortion cost and the robustness cost at the same time. Therefore, RSVRC embedding can improve the robustness of stego image.

\subsection{Robustness performance comparison}
In this part 200 images are selected randomly from JPEG images generated from BossBase-1.0.1-cover database in different quality factors as cover images. The BCH error correction code with block length $n = 127$ and the information length $k=64$ is used to perform error correction coding to the hidden message. The coded message is randomly permutated before embedding into cover images. The results of the experiment are shown in table \ref{tab:2_1}, where the payload is calculated using the length of information bits before error correction coding.
\begin{table} [!htbp]
   \caption{Robust performance of RSVRC and JCRISBE algorithms with different quality factors at different payload size and (127,64) BCH code.}
  \label{tab:2_1}
  \centering
  \begin{tabular}{| p{1.5cm}<{\centering}| p{1.5cm}<{\centering}| p{2cm}<{\centering}| p{2cm}<{\centering}| p{2cm}<{\centering}|p{2cm}<{\centering}|}
    \hline
    Quality factor & Payload size (bpnzac)& Algorithm name & Channel error rate $P_{e}(\%)$ & Information bit error rate $P_{s}(\%)$ & Successful extraction rate $R_{s}(\%) $\\
    \hline
    \multirow{4}{*}{75} &\multirow{2}{*}{0.05} & JCRISBE & 0.0061 & 0.3844 & 97.75 \\ \cline{3-6}
      & & RSVRC &$\mathbf{0.0003}$   & $\mathbf{0.0206}$ & $\mathbf{100}$ \\ \cline{2-6}
      & \multirow{2}{*}{0.1} & JCRISBE & 0.0129 & 0.4075 & 97.25 \\ \cline{3-6}
      & & RSVRC & $\mathbf{0.0005}$ &$\mathbf{0.0149}$  & $\mathbf{100}$\\ \hline
   \multirow{4}{*}{85} &\multirow{2}{*}{0.05} & JCRISBE & 0.0079& 0.3728 & 97.43 \\ \cline{3-6}
      & & RSVRC & $\mathbf{0.0007}$ & $\mathbf{0.0343}$& $\mathbf{100}$\\ \cline{2-6}
      & \multirow{2}{*}{0.1} & JCRISBE & 0.0154 & 0.3623 & 98.12 \\ \cline{3-6}
      & & RSVRC & $\mathbf{0.0011}$ & $\mathbf{0.0240}$ & $\mathbf{100}$\\ \hline
   \multirow{4}{*}{95} &\multirow{2}{*}{0.05} & JCRISBE & 0.5504 & 13.40 & 0\\ \cline{3-6}
      & & RSVRC & $\mathbf{0.4131}$ & $\mathbf{10.15}$ & $\mathbf{0}$\\ \cline{2-6}
      & \multirow{2}{*}{0.1} & JCRISBE & 0.5502& 13.38& 0 \\ \cline{3-6}
      & & RSVRC & $\mathbf{0.4047}$ &$ \mathbf{10.91}$& $\mathbf{0}$\\ \hline
  \end{tabular}
\end{table}

It can be seen from the experimental results that the RSVRC algorithm has a lower channel error rate than the JCRISBE algorithm after recompression at all cases, which shows that the RSVRC algorithm can improve the robustness of the embedding. At the same time, when the recompression quality factor of the channel is 75 and 85, the channel error rate is low, and the channel error rate is relatively high when the quality factor is 95, resulting in the extraction failure using the parameter of error correction code in the experiment. The reason for this phenomenon is that the rounding loss in spatial domain of recompression will cause the unquantized DCT coefficients to fluctuate. Because of TCM operation, the fluctuation would not change the quantized DCT coefficients. After embedding, there is new fluctuation in the unquantized DCT coefficients. When DCT quantization process using a larger quantization step size, the fluctuation of unquantized  DCT coefficients is not large enough to change the results of the rounding operation which generate quantized DCT coefficients. When the quantization step is small, a small amount of changes in the unquantized DCT coefficients will change quantized DCT coefficients. It is difficult to restore the message when the quality factor of recompression is 95. 

\subsection{Security performance comparison}
\begin{figure}
\centering
\includegraphics [width=0.6\textwidth]{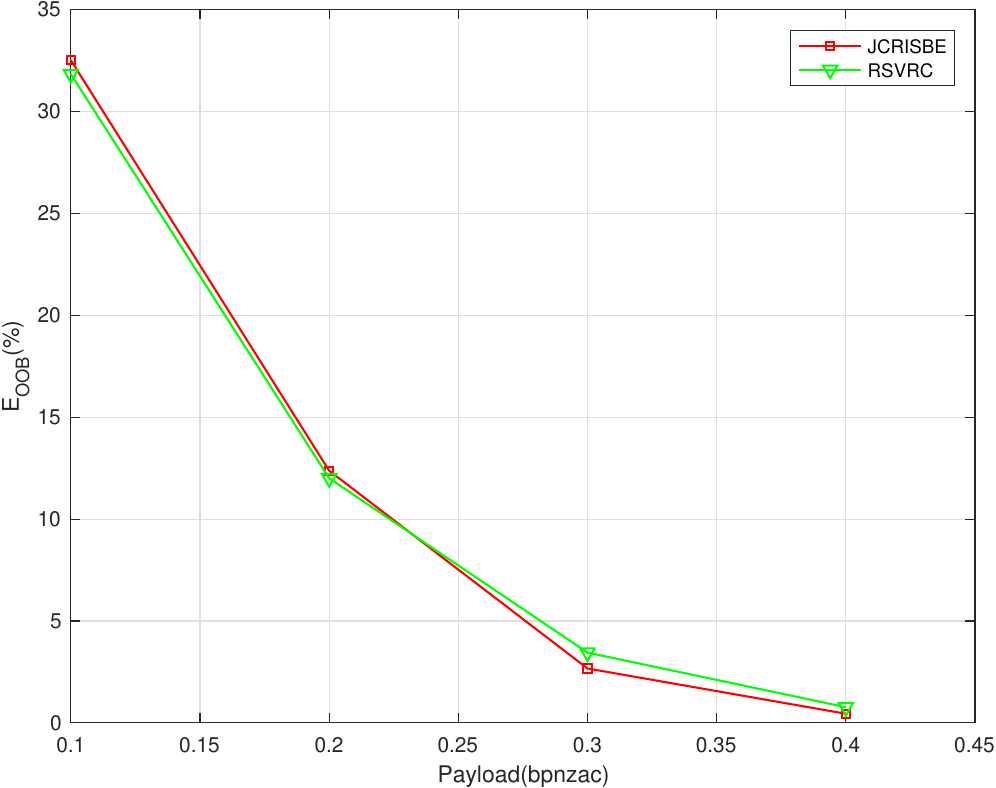}
\caption{Security performance of JCRISBE and RSVRC algorithm.} \label{fig4}
\end{figure}

This section compares the detection error rate of steganalysis in two cases. Firstly, the security of the proposed algorithm and the JCRISBE algorithm under same payload size is compared. Secondly, when embedding same length of information, the proposed algorithm can use error correction coding with higher coding efficiency because of better robustness, so the security of the RSVRC and JCRISBE algorithm with different coding efficiency and same length of message before error encoding is compared. The training set used in the experiment contains 600 pairs of images with and without embedding both after channel matching. The testing set contains 400 pairs of images with and without embedding both after channel matching. The quality factor of recompression channel is 85. The steganalysis algorithm that is used to test the security performance is the GPU-accelerated GFR  \cite {song2015steganalysis,XiaHighly} steganalysis algorithm. The security performance is evaluated using detection error rate $P_{E}$ of the trained classifier, which is defined as:
\begin{equation} \label{pe}
P_{E} = \mathop{\min}_{P_{FA}} \frac{1}{2}(P_{FA}+P_{MD}),
\end{equation}
where $P_{FA}$ is the possibility of misclassifying the cover image as stego image, and $P_{MD}$ is the  possibility of misclassifying the stego image as cover image. The security performance under the different payload is shown in figure \ref{fig4}.

It can be seen from figure \ref{fig4} that when the embedding rate is 0.1 and 0.2, the security of RSVRC algorithm is slightly lower than that of JCRISBE algorithm, and when the embedding rate is 0.3 and 0.4, the security of RSVRC embedded image is slightly higher than that of JCRISBE algorithm.  Although the cost of RSVRC algorithm
not only considers the security but also the robustness, the security of the image after RSVRC steganography is not significantly lower than that of the JCRISBE algorithm, and the robustness of the algorithm has been greatly improved.

The performance of security and robustness of RSVRC and JCRISBE algorithm with different coding efficiency of BCH coding and same length of information before error encoding is show in table \ref{tab:vcrssc}. In the experiment, the length of information bits before error correction coding is 0.1 bpnzac, and the block size of BCH coding $n$ is 127. The quality factor of recompression is 85. It can be seen from table \ref{tab:vcrssc} that when the coding efficiency of BCH coding is improved, even though the information bit error rate of the RSVRC algorithm is increased, but it is still lower than the JCRISBE algorithm.  Due to the reduction in length of embedded message, the interference caused by embedding is reduced, thereby the security is improved. From the detection error rate of the trained classifier, it can be seen that the RSVRC algorithm has better performance in the combination of security and robustness compared with JCRISBE algorithm.
\begin{table} [!htbp]
  \caption{Information bit error rate, detection error rate, and successful extraction rate of RSVRC and JCRISBE algorithms with quality factor 85, payload 0.1 bpnzac and different coding efficiency of BCH code with n = 127.}
  \label{tab:vcrssc}
  \centering
  \footnotesize
  \setlength{\tabcolsep}{4pt}
  \renewcommand{\arraystretch}{1.2}
  \begin{tabular}{ |p{2.1cm}<{\centering}| p{2.1cm}<{\centering}| p{2.1cm}<{\centering}| p{2.1cm}<{\centering}| p{2.1cm}<{\centering}|}
    \hline
    Algorithm name& Coding efficiency $e(\%)$ & Information bit error rate $P_{s}(\%)$ & Detection error  rate $P_{E} (\%)$ &Successful extraction rate $R_{s}(\%)$\\
    \hline
    JCRISBE  &  50.39  &  0.36 & 12.37  & 98.1 \\ \hline
    RSVRC  &  50.39   &$\mathbf{0.02}$   & 12 & 100\\ \hline
    RSVRC  &  $\mathbf{72.44}$  & 0.03 & $\mathbf{22.62}$ &  $\mathbf{100}$ \\ 
    
    \hline
  \end{tabular}
\end{table}

\section{Conclusion}
In order to improve the robustness of the robust steganography algorithm based on the TCM algorithm, this paper proposes a robustness cost to enable STC based embedding algorithm to modify DCT coefficients which have better robustness. Because the robustness of the embedded coefficient is difficult to estimate before embedding and the influence of different modified coefficients interferes with each other on the robustness, the embedding cost is dynamically updated using local recompression simulator. Experimental results show that the algorithm can improve the security and robustness performance using more efficient error coding settings. Since the stego images need to be saved during the embedding process, the algorithm has a high computational complexity. The next research will focus on reducing the complicity of proposed algorithm.

%
%
%
\bibliographystyle{splncs04}
 \bibliography{ref}

\end{document}